\documentclass[11pt]{article}
\usepackage{latexsym,epic,eepic}
\usepackage{graphicx,epsfig}
\begin{document}
\newcommand{\2}{\vspace{0.2 cm}}
\newcommand{\dist}{{\rm dist}}
\newcommand{\diam}{{\rm diam}}
\newcommand{\rad}{{\rm rad}}
\newcommand{\dom}{\mbox{$\rightarrow$}}
\newcommand{\ldom}{\mbox{$\leftarrow$}}
\newcommand{\edom}{\mbox{$\leftrightarrow$}}
\newcommand{\qed}{\hfill$\diamond$}
\newcommand{\pf}{{\bf Proof: }}
\newtheorem{theorem}{Theorem}[section]
\newcommand{\ra}{\rangle}
\newcommand{\la}{\langle}
\newtheorem{lemma}[theorem]{Lemma}
\newtheorem{claim}[theorem]{Claim}
\newtheorem{definition}[theorem]{Definition}
\newtheorem{corollary}[theorem]{Corollary}
\newtheorem{observation}[theorem]{Observation}
\newtheorem{proposition}[theorem]{Proposition}
\newtheorem{conjecture}[theorem]{Conjecture}
\newtheorem{problem}[theorem]{Problem}
\newtheorem{question}[theorem]{Question}
\newtheorem{remark}[theorem]{Remark}
\newcommand{\beq}{\begin{equation}}
\newcommand{\eeq}{\end{equation}}
\newcommand{\UN}{{\rm UN}}
\newcommand{\MiP}{MinHOM($H$) }
\newcommand{\MaP}{MaxHOM($H$) }
\newcommand{\vecc}[1]{\stackrel{\leftrightarrow}{#1}}
\newcommand{\sdom}{\mbox{$\Rightarrow$}}
\newcommand{\nsdom}{\mbox{$\not\Rightarrow$}}
\newcommand{\eps}[2][1]{%
  \scalebox{#1}{\includegraphics{fig/hex.#2}}}
\newcommand{\labcontent}[2]{%
  \vbox{\halign{\hfil##\hfil\cr#1\cr(#2)\cr}}}
\newcommand{\labeps}[3][1]{\labcontent{\eps[#1]{#2}}{#3}}
\newcommand{\fline}[1]{\hbox to \hsize{\hfil#1\hfil}}

\title{Minimum Cost Homomorphisms to Locally Semicomplete and
Quasi-Transitive Digraphs}
\author{A. Gupta\thanks{School of
Computing Science, Simon Fraser University, Burnaby, B.C., Canada,
V5A 1S6, E-mail: arvind@cs.sfu.ca}, G. Gutin\thanks{Department of
Computer Science, Royal Holloway, University of London, Egham, TW20
0EX, UK, E-mail: gutin@cs.rhul.ac.uk}, M.
Karimi\thanks{School of Computing Science, Simon Fraser University,
Burnaby, B.C., Canada, V5A 1S6, E-mail: mmkarimi@cs.sfu.ca}, E.J. Kim\thanks{Department
of Computer Science, Royal Holloway, University of London, Egham,
TW20 0EX, UK, E-mail: e.j.kim@cs.rhul.ac.uk}, A.
Rafiey\thanks{School of Computing Science, Simon Fraser University,
Burnaby, B.C., Canada, V5A 1S6, E-mail: arashr@cs.sfu.ca}}

\date{}

\maketitle

\begin{abstract}
For digraphs $G$ and $H$, a homomorphism of $G$ to $H$ is a mapping
$f:\ V(G)\dom V(H)$ such that $uv\in A(G)$ implies $f(u)f(v)\in
A(H)$. If, moreover, each vertex $u \in V(G)$ is associated with
costs $c_i(u), i \in V(H)$, then the cost of a homomorphism $f$ is
$\sum_{u\in V(G)}c_{f(u)}(u)$. For each fixed digraph $H$, the
minimum cost homomorphism problem for $H$, denoted MinHOM($H$), can
be formulated as follows: Given an input digraph $G$, together with
costs $c_i(u)$, $u\in V(G)$, $i\in V(H)$, decide whether there
exists a homomorphism of $G$ to $H$ and, if one exists, to find one
of minimum cost. Minimum cost homomorphism problems encompass (or
are related to) many well studied optimization problems such as the
minimum cost chromatic partition and repair analysis problems. We
focus on the minimum cost homomorphism problem for locally
semicomplete digraphs and quasi-transitive digraphs which are two
well-known generalizations of tournaments. Using graph-theoretic
characterization results for the two digraph classes, we obtain a
full dichotomy classification of the complexity of minimum cost
homomorphism problems for both classes.

\end{abstract}

\section{Introduction}\label{introsec}

The minimum cost homomorphism problem was introduced in \cite{gutinDAMlora}, where it was
motivated by a real-world problem in defense logistics. In general,
the problem appears to offer a natural and practical way to model
many optimization problems. Special cases include the homomorphism
problem, the list homomorphism problem \cite{hell2003,hell2004} and
the optimum cost chromatic partition problem
\cite{halld2001,jansenJA34,jiangGT32} (which itself has a number of
well-studied special cases and applications
\cite{kroon1997,supowitCAD6}).

For digraphs $G$ and $H$, a mapping $f:\ V(G)\dom V(H)$ is a {\em
homomorphism of $G$ to $H$} if $f(u)f(v)$ is an arc of $H$ whenever
$uv$ is an arc of $G$. In the {\em homomorphism problem}, given a
graph $H$, for an input graph $G$ we wish to decide whether there is
a homomorphism of $G$ to $H$. In the {\em list homomorphism
problem}, our input apart from $G$ consists of sets $L(u)$, $u\in
V(G)$, of vertices of $H$, and we wish to decide whether there is a
homomorphism $f$ of $G$ to $H$ such that $f(u)\in L(u)$ for each
$u\in V(G)$. In the {\em minimum cost homomorphism problem} we fix
$H$ as before, our inputs are a graph $G$ and costs $c_i(u),\ u\in
V(G),\ i \in V(H)$ of mapping $u$ to $i$, and we wish to check
whether there exists a homomorphism of $G$ to $H$ and if it does
exist, we wish to obtain one of minimum cost, where the cost of a
homomorphism $f$ is $\sum_{u\in V(G)}c_{f(u)}(u)$. The homomorphism,
list homomorphism, and minimum cost homomorphism problems are
denoted by HOM($H$), ListHOM($H$) and MinHOM($H$), respectively. If
the graph $H$ is {\em symmetric} (each $uv \in A(H)$ implies $vu \in
A(H)$), we may view $H$ as an undirected graph. This way, we may
view the problem MinHOM($H$) as also a problem for undirected
graphs. For {\em further terminology and notation} see the next section, 
where we define several terms used in the rest of this section.

Our interest is in obtaining dichotomies: given a problem such as
HOM($H$), we would like to find a class of digraphs $\cal H$ such
that if $H\in {\cal H}$, then the problem is polynomial-time
solvable and if $H\notin {\cal H}$, then the problem is NP-complete.
For instance, in the case of undirected graphs it is well-known that
HOM($H$) is polynomial-time solvable when $H$ is bipartite or has a
loop, and NP-complete otherwise \cite{hellJCT48}. 

For undirected graphs $H$, a dichotomy classification for the
problem MinHOM($H$) has been provided in \cite{mincostungraph}. (For
ListHOM$(H)$, consult \cite{pavol}.) Since \cite{mincostungraph}
interest has shifted to directed graphs. The first studies
\cite{gutinDAM,gutinDO,yeo} focused on loopless digraphs and
dichotomies have been obtained for semicomplete digraphs and
semicomplete multipartite digraphs (we define these and other
classes of digraphs in the next section). More recently,
\cite{gregorykim2} initiated the study of digraphs with loops
allowed; and, in particular, of reflexive digraphs, where each
vertex has a loop. While \cite{gregorykim} gave a dichotomy for
semicomplete digraphs with possible loops, \cite{hellrefdig}
obtained a dichotomy for all reflexive digraphs. (Partial results on
ListHOM$(H)$ for digraphs can be found in \cite{Hellpowerdigraph,
feder, Hellpartitioning, federmanuscript,Federcycle,helltree,Zhou}.)

Along with semicomplete digraphs and semicomplete multipartite
digraphs, locally semicomplete digraphs and quasi-transitive
digraphs are the most studied families of generalizations of
tournaments \cite{bang2000}. Thus, it is a natural problem to obtain
dichotomies for locally semicomplete digraphs and quasi-transitive
digraphs and we solve this problem in the present paper. Like with
semicomplete digraphs and semicomplete multipartite digraphs,
structural properties of locally semicomplete digraphs and
quasi-transitive digraphs play key role in proving the dichotomies.
Unlike for semicomplete digraphs and semicomplete multipartite
digraphs, we also use structural properties of a family of
undirected graphs. We hope that the study of well-known classes of
digraphs will eventually allow us to conjecture and prove a full
dichotomy for loopless digraphs.

In this paper we prove the following two dichotomies:

\begin{theorem}\label{localysemicomplete}
Let $H$ be a locally semicomplete digraph. MinHom($H$) is
polynomial-time solvable if every connectivity component of $H$ is
either acyclic or a directed cycle $\overrightarrow{C_k}$, $k\ge 2$.
Otherwise, MinHom($H$) is NP-hard.
\end{theorem}

\begin{figure}[t]
\begin{center}
\epsfig{file=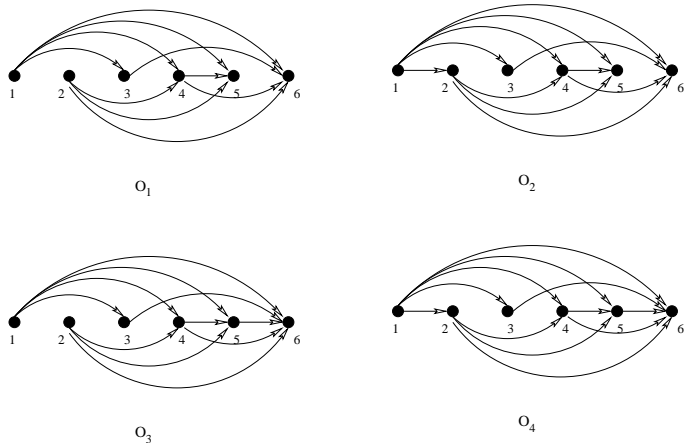, width=9cm}
\end{center}
\caption{The obstructions $O_i$ with $i=1,2,3,4 $} \label{O1}
\end{figure}

\begin{theorem}\label{quasi}
Let $H$ be a quasi-transitive digraph. MinHom($H$) is
polynomial-time solvable if every connectivity component $H'$ of $H$
is either $\overrightarrow{C_2}$ or an extension of
$\overrightarrow{C_3}$ or acyclic, $B(H')$ is a proper interval
bigraph and $H'$ does not contain $O_i$ with $i=1,2,3,4$ as an
induced subgraph (the digraphs $O_i$ are defined as in Figure
\ref{O1}). Otherwise, MinHom($H$) is NP-hard.
\end{theorem}

In fact, it is easy to see that it suffices to prove Theorems
\ref{localysemicomplete} and \ref{quasi} only for connected digraphs
$H$; for a short proof, see \cite{mincostungraph}. The rest of the
paper is devoted to proving the two theorems for the case of
connected $H$. In Section \ref{termnotsec} we provide further
terminology and notation and formulate a characterization of proper
interval bigraphs that we use later. In Section \ref{psec} we prove
the polynomial-time solvability parts of the two theorems. While the
proof of the polynomial-time solvability part of Theorem
\ref{localysemicomplete} is relatively easy, this part of Theorem
\ref{quasi} is quite technical and lengthy. In Section \ref{psec} we
prove the NP-completeness parts of the two theorems. There we use
several known results and prove some new ones.

\section{Further Terminology and Notation}\label{termnotsec}

In our terminology and notation, we follow \cite{bang2000}. From now
on, all digraphs are loopless and do not have parallel arcs. A
digraph $D$ is {\em semicomplete} if, for each pair $x, y$ of
distinct vertices either $x$ dominates $y$ or $y$ dominates $x$ or
both. A digraph $D$ obtained from a complete $k$-partite
(undirected) graph $G$ by replacing every edge $xy$ of $G$ with arc
$xy$, arc $yx$, or both, is called a {\em semicomplete k-partite
digraph} (or, semicomplete multipartite digraph when $k$ is
immaterial). A digraph $D$ is {\em locally semicomplete} if for
every vertex $x$ of $D$, the in-neighbors of $x$ induce a
semicomplete digraph and the out-neighbors of $x$ also induce a
semicomplete digraph. A digraph $D$ is {\em transitive} if, for
every pair of arcs $xy$ and $yz$ in $D$ such that $x \neq z$, the
arc $xz$ is also in $D$. Sometimes, we will deal with {\em
transitive oriented graphs}, i.e. transitive digraphs with no cycle
of length two. A digraph $D$ is {\em quasi-transitive} if, for every
triple $x,y,z$ of distinct vertices of $D$ such that $xy$ and $yz$
are arcs of $D$, there is at least one arc between $x$ and $z$.
Clearly, a transitive digraph is also quasi-transitive. Notice that
a semicomplete digraph is both quasi-transitive and locally
semicomplete.

An ($x,y$)-path in a digraph $D$ is a directed path from $x$ to $y$.
A digraph $D$ is {\em strongly connected} (or, just, {\em strong})
if, for every pair $x,y$ of distinct vertices in $D$, there exist an
($x,y$)-path and a ($y,x$)-path. A {\em strong component} of a
digraph $D$ is a maximal induced subgraph of $D$ which is strong. If
$D_1,\ldots,D_t$ are the strong components of $D$, then clearly
$V(D_1) \cup \ldots \cup V(D_t) = V(D)$ (recall that a digraph with
only one vertex is strong). Moreover, we must have $V(D_i) \cap
V(D_j) = \emptyset$ for every $i \neq j$ as otherwise all the
vertices $V(D_i) \cup V(D_j)$ are reachable from each other,
implying that the vertices of $V(D_i) \cup V(D_j)$ belong to the
same strong component of $D$.

Let $D$ be any digraph. If $xy \in A(D)$, we say $x$ dominates $y$
or $y$ is dominated by $x$, and denote by $x \dom y$. An arc $xy \in
A(D)$ is {\em symmetric} if $yx \in A(D)$. For sets $X,Y \subset
V(D)$, $X \dom Y$ means that $x \dom y$ for each $x \in X, y \in Y$. 
We denote by $B(D)$
the bipartite graph obtained from $D$ as follows. Each vertex $v$ of
$D$ gives rise to two vertices of $B(D)$ - a {\em white} vertex $v'$
and a {\em black} vertex $v''$; each arc $vw$ of $D$ gives rise to
an edge $v'w''$ of $B(D)$. Note that if $D$ is an irreflexive
digraph, then all edges $v'v''$ are absent in $B(D)$. The {\em
converse} of $D$ is the digraph obtained from $D$ by reversing the
directions of all arcs. A digraph $H$ is an {\em extension} of $D$
if $H$ can be obtained from $D$ by replacing every vertex $x$ of $D$
with a set $S_x$ of independent vertices such that if $xy \in A(D)$
then $uv \in A(H)$ for each $u \in S_x, v \in S_y$. A {\em
tournament} is a semicomplete digraph which does not have any
symmetric arc. An acyclic tournament on $p$ vertices is denoted by
$TT_p$ and called a {\em transitive tournament}. The vertices of a
transitive tournament $TT_p$ can be labeled $1,2\ldots, p$ such that
$ij \in A(TT_p)$ if and only if $1 \leq i < j \leq p$. By $TT^-_p$
($p \geq 2$), we denote $TT_p$ without the arc $1p$.

We say that a bipartite graph $H$ (with a fixed bipartition into
white and black vertices) is a {\em proper interval bigraph} if
there are two inclusion-free families of intervals $I_v$, for all
white vertices $v$, and $J_w$ for all black vertices $w$, such that
$vw \in E(H)$ if and only if $I_v$ intersects $J_w$. By this
definition proper interval bigraphs are irreflexive and bipartite. A
combinatorial characterization (in terms of forbidden induced
subgraphs) of proper interval bigraphs is given in
\cite{hellhuang2004}: $H$ is a proper interval bigraph if and only
if it does not contain an induced cycle $C_{2k}$, with $k \geq 3$,
or an induced biclaw, binet, or bitent, as given in Figure
\ref{bipartitenettentclaw}.

A linear ordering $<$ of $V(H)$ is a {\em Min-Max ordering} if $i<j,
s<r$ and $ir, js \in A(H)$ imply that $is \in A(H)$ and $jr \in
A(H)$. For a bipartite graph $H$ (with a fixed bipartition into
white and black vertices), it is easy to see that $<$ is a Min-Max
ordering if and only if $<$ restricted to the white vertices, and
$<$ restricted to the black vertices satisfy the condition of
Min-Max orderings, i.e., $i<j$ for white vertices, and $s<r$ for
black vertices, and $ir, js \in A(H)$, imply that $is \in A(H)$ and
$jr \in A(H)$). A {\em bipartite Min-Max ordering} is an ordering
$<$ specified just for white and for black vertices.

The following lemma exhibits that a proper interval bigraph always
admits a bipartite Min-Max ordering.

\begin{lemma} \cite{mincostungraph}\label{pibminmaxordering}
A bipartite graph $G$ is a proper interval bigraph if and only if
$G$ admits a bipartite Min-Max ordering.
\end{lemma}

It is known that if $H$ admits a Min-Max ordering, then the problem
MinHOM($H$) is polynomial-time solvable \cite{gutinDAM}, see also
\cite{cohenJAIR22,khana}; however, there are digraphs $H$ for which
MinHOM($H$) is polynomial-time solvable, but $H$ has no Min-Max orderings
\cite{gutinDO}.

\begin{figure}
\unitlength 0.330mm \linethickness{0.4pt} \noindent
\begin{picture}(   120.00,   100.00)
\put(    60.00,    90.00){\circle{12.0}} \put(  60.000,
90.000){\makebox(0,0){{\scriptsize $x_1$}}} \put(   110.00,
10.00){\circle{12.0}} \put( 110.000,
10.000){\makebox(0,0){{\scriptsize $x_2$}}} \put(    10.00,
10.00){\circle{12.0}} \put(  10.000,
10.000){\makebox(0,0){{\scriptsize $x_3$}}} \put(    60.00,
40.00){\circle{12.0}} \put(  60.000,
40.000){\makebox(0,0){{\scriptsize $x_4$}}} \put(    60.00,
65.00){\circle{12.0}} \put(  60.000,
65.000){\makebox(0,0){{\scriptsize $y_1$}}} \put(    85.00,
25.00){\circle{12.0}} \put(  85.000,
25.000){\makebox(0,0){{\scriptsize $y_2$}}} \put(    35.00,
25.00){\circle{12.0}} \put(  35.000,
25.000){\makebox(0,0){{\scriptsize $y_3$}}} \drawline(  60.000,
84.000)(  60.000,  71.000) \drawline(  60.000,  59.000)(  60.000,
46.000) \drawline( 104.855,  13.087)(  90.145,  21.913) \drawline(
79.855,  28.087)(  65.145,  36.913) \drawline(  15.145,  13.087)(
29.855,  21.913) \drawline(  40.145,  28.087)(  54.855,  36.913)
\put(60,-10){\makebox(0,0){{\footnotesize (a)}}}
\end{picture} \hspace{-0.01cm}
\linethickness{0.4pt}
\begin{picture}(   120.00,   100.00)
\put(   110.00,    90.00){\circle{12.0}} \put( 110.000,
90.000){\makebox(0,0){{\scriptsize $x_1$}}} \put(    10.00,
10.00){\circle{12.0}} \put(  10.000,
10.000){\makebox(0,0){{\scriptsize $x_2$}}} \put(   110.00,
10.00){\circle{12.0}} \put( 110.000,
10.000){\makebox(0,0){{\scriptsize $x_3$}}} \put(    60.00,
50.00){\circle{12.0}} \put(  60.000,
50.000){\makebox(0,0){{\scriptsize $x_4$}}} \put(   110.00,
50.00){\circle{12.0}} \put( 110.000,
50.000){\makebox(0,0){{\scriptsize $y_1$}}} \put(    60.00,
10.00){\circle{12.0}} \put(  60.000,
10.000){\makebox(0,0){{\scriptsize $y_2$}}} \put(    20.00,
80.00){\circle{12.0}} \put(  20.000,
80.000){\makebox(0,0){{\scriptsize $y_3$}}} \drawline( 110.000,
84.000)( 110.000,  56.000) \drawline( 110.000,  44.000)( 110.000,
16.000) \drawline( 104.000,  10.000)(  66.000,  10.000) \drawline(
60.000,  16.000)(  60.000,  44.000) \drawline(  66.000,  50.000)(
104.000,  50.000) \drawline(  16.000,  10.000)(  54.000,  10.000)
\drawline(  24.800,  76.400)(  55.200,  53.600)
\put(60,-10){\makebox(0,0){{\footnotesize (b)}}}
\end{picture}  \hspace{-0.01cm}
\linethickness{0.4pt}
\begin{picture}(   120.00,   100.00)
\put(    60.00,    10.00){\circle{12.0}} \put(  60.000,
10.000){\makebox(0,0){{\scriptsize $x_1$}}} \put(    10.00,
50.00){\circle{12.0}} \put(  10.000,
50.000){\makebox(0,0){{\scriptsize $x_2$}}} \put(    60.00,
90.00){\circle{12.0}} \put(  60.000,
90.000){\makebox(0,0){{\scriptsize $x_3$}}} \put(   110.00,
50.00){\circle{12.0}} \put( 110.000,
50.000){\makebox(0,0){{\scriptsize $x_4$}}} \put(    60.00,
50.00){\circle{12.0}} \put(  60.000,
50.000){\makebox(0,0){{\scriptsize $y_1$}}} \put(   110.00,
10.00){\circle{12.0}} \put( 110.000,
10.000){\makebox(0,0){{\scriptsize $y_2$}}} \put(    10.00,
10.00){\circle{12.0}} \put(  10.000,
10.000){\makebox(0,0){{\scriptsize $y_3$}}} \drawline(  60.000,
16.000)(  60.000,  44.000) \drawline(  60.000,  56.000)(  60.000,
84.000) \drawline(  16.000,  50.000)(  54.000,  50.000) \drawline(
10.000,  44.000)(  10.000,  16.000) \drawline( 104.000,  50.000)(
66.000,  50.000) \drawline( 110.000,  44.000)( 110.000,  16.000)
\drawline(  54.000,  10.000)(  16.000,  10.000) \drawline( 66.000,
10.000)( 104.000,  10.000) \put(60,-10){\makebox(0,0){{\footnotesize
(c)}}}
\end{picture}

\mbox{ }

\caption{A biclaw  (a), a binet (b) and a bitent (c).}
\end{figure}
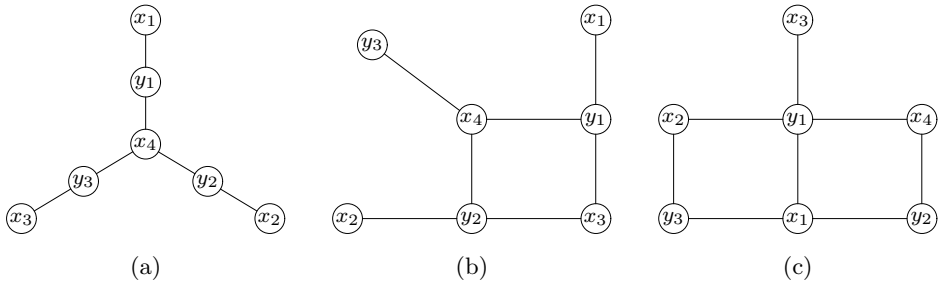  \label{bipartitenettentclaw}

\section{Polynomial cases }\label{psec}
The most basic properties of strong components of a connected
non-strong locally semicomplete digraph are given in the following
result, due to Bang-Jensen \cite{Bangjensenlocal}.

\begin{theorem} \cite{Bangjensenlocal}\label{localysemicomplete0}
Let $H$ be a connected locally semicomplete digraph that is not
strong. Then the following holds for $H$.

\begin{description}
\item{(a)} If $A$ and $B$ are distinct strong components of $H$ with
at least one arc between them, then either $A \dom B$ or $B \dom A$.

\item{(b)} If $A$ and $B$ are strong components of $H$, such that $A
 \dom B$, then $A$ and $B$ are semicomplete digraphs.

\item{(c)} The strong components of $H$ can be ordered in a unique way
$H_1,H_2,\ldots, H_p$ such that there are no arcs from $H_j$ to
$H_i$ for $j > i$, and $H_i$ dominates $H_{i+1}$ for $i=1,2,\ldots,
p-1$.
\end{description}
\end{theorem}

\begin{theorem}\label{polylocalysemicomplete}
Let $H$ be a connected locally semicomplete digraph. MinHom($H$) is
polynomial time solvable if $H$ is either acyclic or a directed
cycle $\overrightarrow{C_k}$, $k \geq 2$.
\end{theorem}

\pf We already know that MinHom($H$) is polynomial time solvable
when $H$ is a directed cycle, see \cite{gutinDAM}. Assume that $H$
is a locally semicomplete digraph which is acyclic. Then $H$ is
non-strong and every strong component of $H$ is a single vertex.
Hence we know from Part (c) of Theorem \ref{localysemicomplete0}
that the vertices of $H$ can be ordered in a unique way $1,2,\ldots,
p$ such that there are no arcs from $j$ to $i$ for $j>i$, and $i$
dominates $i+1$ for $i=1,2,\ldots, p-1$. We claim that this ordering
is a Min-Max ordering and thus, MinHom($H$) is polynomial time
solvable.

Choose any two arcs $ir, js \in A(H)$ with $i<j$, $s<r$. Since all
arcs are oriented forwardly with respect to the ordering, we have $i
< j < s < r$. Also, there is a path $i, i+1,\ldots, j, \ldots ,s,
s+1, \ldots, r$ from $i$ to $r$ in $H$ due to the ordering property.
As vertex $i$ dominates both $i+1$ and $r$, there is an arc between
$i+1$ and $r$, which should be oriented from $i+1$ to $r$. By
induction, vertex $r$ is dominated by all vertices $i, \ldots, r-1$
on the path. This indicates that we have an arc $jr \in A(H)$.
Following a similar argument, we conclude that there is an arc $is
\in A(H)$. This proves that the ordering is a Min-Max ordering. \qed



\begin{theorem}\label{polyquasi}
Let $H$ be a connected quasi-transitive digraph. Then MinHom($H$) is
polyonomial time solvable if either
\begin{itemize}
\item $H$ is $\overrightarrow{C_2}$ or $H$ is an extension of
$\overrightarrow{C_3}$, or
\item $H$ is acyclic, $B(H)$ is a proper interval bigraph and $H$
does not contain $O_i$ with $i=1,2,3,4 $ as an induced subdigraph.
(See Figure \ref{O1}.)
\end{itemize}
\end{theorem}

\pf It has been proved in \cite{gutinDAM} that MinHom($H$) is
polynomial time solvable when $H$ is a directed cycle. The case for
$H$ being $\overrightarrow{C_2}$ or $\overrightarrow{C_3}$ follows
immediately.

Now assume that $H$ is acyclic. Then, it is straightforward to check
that $H$ is exactly a transitive oriented graph $T$. We will show
that a bipartite Min-Max ordering of $B(T)$ can be transformed to
produce a Min-Max ordering of $T$. Recall that whenever $B(T)$ is a
proper interval bigraph it has a bipartite Min-Max ordering due to
Lemma \ref{pibminmaxordering}.

Suppose $<$ is a bipartite Min-Max ordering of $B(T)$. A pair $x, y$
of vertices of $T$ is {\em proper for $<$} if $x' < y'$ if and only
if $x'' < y''$ in $B(T)$. We say a bipartite Min-Max ordering $<$ is
{\em proper} if all pairs $x, y$ of $T$ are proper for $<$. If $<$
is a proper bipartite Min-Max ordering, then we can define a
corresponding ordering $\prec$ on the vertices of $T$, where $x
\prec y$ if and only if $x' < y'$ (which happens if and only if $x''
< y''$). It is easy to check that $\prec$ is now a Min-Max ordering
of $T$.

Assume, on the other hand, that the bipartite Min-Max ordering $<$
on $B(T)$ is not proper. That is, there are vertices $x', y'$ such
that $x' < y'$ and $y'' < x''$. Suppose that for every pair of
vertices $c''$ and $d''$ such that $d''< c''$ and $x'd'',y'c'' \in
E(B(T))$, we have both $x'c''$ and $y'd''$ in $E(B(T))$. Then we can
exchange the positions of $x'$ and $y'$ in $<$ while perserving the
Min-Max property. Furthermore, it can be checked that this exchange
strictly increases the number of proper pairs in $H$: if a proper
pair turns into an improper pair or vice versa by this exchange,
then one of the two vertices should be $x$ or $y$. Clearly the
improper pair consisting of $x$ and $y$ is turned into a new proper
pair. Suppose that vertex $w$ constitues a pair with $x$ or $y$
which is possibly affected by the exchange. Observe that we have
$x'<w'<y'$ or $y''<w''<x''$. When $w$ lies between $x$ and $y$ in
both partite sets in $B(T)$, the improper pairs $(w,x)$, $(w,y)$ are
transformed to proper pairs by the exchange of $x'$ and $y'$. When
$x'<w'<y'$ and $w''$ is not between $x''$ and $y''$, there is a
newly created proper pair and improper pair respectively, which
compensate the effect of each other in the number of proper pairs in
$H$. Similarly, there is no change in the number of proper pairs of
the form $(w,x)$ or $(w,y)$ when $y''<w''<x''$ and $w'$ is not
between $x'$ and $y'$. Hence, the exchange increases the number of
proper pairs at least by one.

Analogously, we can exchange the positions of $x''$ and $y''$ in $<$
if for every pair of vertices $a'$ and $b'$ such that $b'< a'$ and
$a'x'',b'y'' \in E(B(T))$, we have both $a'y''$ and $b'x''$ in
$E(B(T))$. This exchange does not affect the Min-Max ordering of
$B(T)$ and strictly increases the number of proper pairs as well.

In the remaining part, we will show that we can always exchange the
positions of $x'$ and $y'$ or the positions of $x''$ and $y''$ in
$<$ whenever we have an improper pair $x$, $y$ and $<$ is a Min-Max
ordering of $B(T)$.

Suppose, to the contrary, that we performed the above exchange for
every improper pair as far as possible and still the Min-Max
ordering is not proper. Then, there must be an improper pair $x$ and
$y$ with $x'<y'$, $y''<x''$ in $<$ which satisfies the following
condition: 1) there exist vertices $c''$ and $d''$, $d''< c''$ such
that $x'd'', y'c'' \in E(B(T))$ and at least one of $y'd''$ and
$x'c''$ is missing in $B(T)$. 2) there exist vertices $a'$ and $b'$,
$b'<a'$ such that $b'y'',a'x'' \in E(B(T))$ and at least one of
$b'x''$ and $a'y''$ is missing in $B(T)$.

Notice that $a$, $d$ and $x$ are distinct vertices in $T$ since
otherwise, the edges $a'x''$ and $x'd''$ induce
$\overrightarrow{C_2}$ or a loop in $T$. With the same argument
$b$,$c$ and $y$ are distict vertices in $T$. On the other hand, by
transitivity of $T$, the edges $a'x''$ and $x'd''$ imply the
existence of edge $a'd''$ in $E(B(T))$. Similarly, there is an edge
$b'c''$ in $E(B(T))$. Note that we do not have $x'x''$ and $y'y''$
in $E(B(T))$ as $T$ is irreflexive.

We will consider cases according to the positions of $a', b', c'',
d''$ in the ordering $<$. We remark the two edges $b'y''$ and
$y'c''$ cannot cross each other. That is, they either satisfy
$b'<y'$ and $y''<c''$, or $y'<b'$ and $c''<y''$, since otherwise
there should be an edge $y'y''$ by the Min-Max property, which is a
contradiction. Similarly, the two edges $a'x''$ and $x'd''$ cannot
cross each other, since otherwise there should be an edge $x'x ''$
by the Min-Max property, which is a contradiction. Hence we have
either $x'<a'$ and $d''<x''$, or $a'<x'$ and $x''<d''$.

When $y'<b'$ and $c''<y''$, the positions of all vertices are
determined immediately so that we have $x'<y'<b'<a'$ and
$d''<c''<y''<x''$. On the other hand, when $b'<y'$ and $y''<c''$ we
can place the edges $a'x''$ and $x'd''$ in two ways, namely to
satisfy either $x'<a'$ and $d''<x''$, or $a'<x'$ and $x''<d''$ due
to the argument in the above paragraph. In the latter case, however,
the positions of all vertices are determined as well and this is
just a converse of the case when $y'<b'$ and $c''<y''$. Therefore we
may assume that $x'<a'$ and $d''<x''$ whenever $b'<y'$ and
$y''<c''$.

{\bf CASE 1} $b'<y'$ and $y''<c''$ ($x'<a'$ and $d''<x''$)

There are following cases to consider. We show that in every case we
have a contradiction.

{\bf Case 1-1} $y' < a'$ and $d'' < y''$

The two edges $a'd'', y'c'' \in E(B(H))$ imply the existence $y'd''
\in E(B(T))$ by the Min-Max property. The edge $y'd''$, however,
together with $b'y'' \in E(B(H))$ enforce the edge $y'y'' \in
E(B(H))$, which is a contradiction.

{\bf Case 1-2} $y' \leq a'$ and $y'' \leq d''$($ < x''$)

Case 1-2-1: $b' < x'$. We know that $a'd'' \in E(B(T))$. We can
easily see $y'd''\in E(B(T))$ since $<$ is a Min-Max ordering . (
Note that $y'c'', a'd'' \in E(B(T))$). By the taking of two vertices
$c'', d''$, the existence of $y'd'' \in E(B(T))$ enforces $x'c''
\not\in E(B(T))$. On the other hand, however, we should have the
edge $x'c'' \in E(B(H))$ due to edges $b'c'', x'd'' \in E(B(H))$ and
the Min-Max property, a contradiction.

Case 1-2-2: $x' \leq b' < y'$. If $x' = b'$ or $y''=d''$ then $x'y''
\in E(B(T))$ since $b'y'' \in E(B(T))$ and $x'd'' \in E(B(T))$. If
$x' < b'$ and $y''<d''$ it is easy to see that we have $x'y'' \in
E(B(T))$ by the Min-Max property. (Note that $b'y'', x'd'' \in
E(B(T))$). With $a'x'',x'y'' \in E(B(H))$, the transitivity of $T$
implies $a'y''\in E(B(T))$. However, this is a contradiction since
we have $y'y'' \not\in E(B(H))$ by the Min-Max property and $y'c'',
a'y'' \in E(B(H))$.

{\bf Case 1-3} ($x'< )a' \leq y'$ and $y'' \leq d''( < x''$)

Case 1-3-1: $x'' < c''$. We will show that we cannot avoid having
the edge $x'c''\in E(B(T))$. Once this is the case, the two edges
$x'c''$ and $a'x''$ imply the existence of edge $x'x''\in E(B(T))$,
which is a contradiction.

When $x' \leq b'$ we again easily observe that $x'y'' \in E(B(T))$
and thus, $x'c'' \in E(B(T))$ for $T$ is transitive and $x'y'',
y'c'' \in E(B(T))$. On the other hand, when $b' < x'$ we have
$x'c''\in E(B(T))$ again by the Min-Max property and the two edges
$b'c'', x'd'' \in E(B(T))$.

Case 1-3-2: $c'' \leq x''$.  We again easily observe that $y'x'' \in
E(B(T))$ by the Min-Max property and the two edges $y'c'', a'x'' \in
E(B(T))$.

When $x' \leq b'$, the Min-Max property implies $x'y'' \in E(B(T))$.
Since $T$ does not contain $\overrightarrow{C_2}$ as an induced
subgraph, this is a contradiction.

When $b' < x'$. It is again implied that $b'x'' \in E(B(T))$ as $T$
is transitive and $b'y'', y'x'' \in E(B(T))$. The two edges $b'x'',
x'd'' \in E(B(H))$ enforce the existence $x'x'' \in E(B(T))$ by the
Min-Max peoperty, which is a contradiction.

{\bf Case 1-4} ($x'< )a' \leq y'$ and $d'' < y''$

We will show that we cannot avoid having the edge $b'x''\in
E(B(T))$. Once this is the case, by the taking of two vertices $a',
b'$, the existence of $b'x'' \in E(B(T))$ enforces $a'y'' \not\in
E(B(T))$. On the other hand, however, we should have the edge $a'y''
\in E(B(H))$ due to edges $a'd'', b'y'' \in E(B(H))$ and the Min-Max
property, a contradiction.

When $x''=c''$, we trivially have $b'x''\in E(B(T))$. When
$x''<c''$, the Min-Max property and the two edges $b'c'',a'x''\in
E(B(T))$ implies $b'x''\in E(B(T))$. When $x''>c''$, the Min-Max
property and the two edges $a'x'',y'c''\in E(B(T))$ implies
$y'x''\in E(B(T))$. For $b'y'',y'x''\in E(B(T))$, we again have
$b'x''\in E(B(T))$ by the transitiviety of $T$. This completes the
argument.

{\bf CASE 2} $y'<b'$ and $c''<y''$

We now prove $T$ has one of $O_i$ with $i = 1,2,3,4$ as an induced
subgraph. Remember that $x' < y' < b' < a'$ and $d'' < c'' < y'' <
x''$. On the other hand, as $T$ is transitive we have $a'd'', b'c''
\in E(B(T))$. Since $<$ is a bipartite Min-Max ordering, $\{a'x'',
a'y'', a'c'', a'd'', b'y'', b'c'', b'd'', y'c'', y'd'', x'd''\}
\subset E(B(T))$. Now by the taking of $a,b$ and $c,d$ we have
$b'x'', x'c'' \not\in E(B(T))$; hence $y'x'', x'y'' \not\in E(B(T))$
as $<$ is a bipartite Min-Max ordering. It is easy to see from the
set of edges existing in $B(T)$ that $a,b,x,y,c,d$ are distinct
vertices in $T$ . Let us define $T'= T[\{a,b,x,y,c,d\}]$. As $T'$ is
acyclic we do not have symmetric arcs in $T'$.

From $E(B(T'))$, we have $\{ax, ay, ac, ad, by, bc, bd, xd, yc, yd\}
\subset A(T')$ and $xy, yx, bx, xc \not\in A(T')$. We can easily see
that $xb \not\in A(T')$, since otherwise from $xb, by \in A(T')$ and
the transitivity of $T'$ we should have $xy \in A(T')$, a
contradiction. With the same argument we will see that $ba, cx, dc
\not\in A(T')$. Therefore we can only add a subset of $S= \{ab,
cd\}$ to the previous arc subset of $T'$ mentioned above each of
which makes $T'$ to be isomorphic to one of $O_i$ with $i=1,2,3,4$
with the mapping $g$ where $g(a)=1, g(b)= 2, g(x)=3, g(y)=4, g(c)=5,
g(d)=6$. \qed

\section{NP-Completeness}\label{NPsec}

We begin this section with a few simple observations. The first one
is easily proved by setting up a natural polynomial time reduction
from MinHOM($B(H)$) to MinHOM$(H)$ \cite{gregorykim2}.

\begin{proposition}\cite{gregorykim2} \label{reduction1}
If MinHOM($B(H)$) is NP-hard, then $MinHOM(H)$ is also NP-hard. \qed
\end{proposition}

The next observation is folklore, and proved by obvious reduction,
cf. \cite{gregorykim}.

\begin{proposition}\label{subdigraph}
Let $H'$ be an induced subgraph of the digraph $H$. If MinHOM($H'$)
is NP-hard, then MinHOM($H$) is NP-hard. \qed
\end{proposition}

The following lemma is the NP-hardness part of the main result in
\cite{gutinDAM}.

\begin{lemma}\cite{gutinDAM}\label{NPsemicomplete}
Let $H$ be a semicomplete digraph containing a cycle and let $H
\not\in \{\overrightarrow{C_2}, \overrightarrow{C_3}\}$. Then
MinHom($H$) is NP-hard.
\end{lemma}

We need two more lemmas for our classification.

\begin{figure}[t]
\begin{center}
\epsfig{file=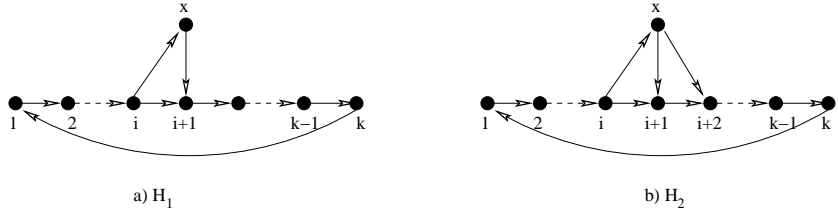, width=11cm}
\end{center}
\caption{$H_1$ and $H_2$.} \label{H1andH2}
\end{figure}

\begin{lemma}\label{h1}
Let $H'_1$ be a digraph obtained from $\overrightarrow{C_k}=12\ldots
k1, k \geq 2$, by adding an extra vertex $k+1$ such that $i \dom
k+1$ and $k+1 \dom i+1$, where $i, i+1$ are two consecutive vertices
in $\overrightarrow{C_k}$. Let $H_1$ be $H'_1$ or its converse. Then
MinHom($H_1$) is NP-hard. (See Figure \ref{H1andH2}.)
\end{lemma}

\pf Without loss of generality, we may assume that
$V(H_1)=\{1,\ldots ,k+1\}$, $123\ldots k$ is a cycle of length $k$,
and the vertex $k+1$ is dominated by $k$ and dominates 1.

We will construct a polynomial time reduction from the maximum
independent set problem to MinHOM($H_1$). Let $G$ be an arbitrary
undirected graph. We replace every edge $uv \in E(G)$ by the digraph
$D_{uv}$ defined as follows:

$V(D_{uv})=\{c_1,c_2,\ldots ,c_{k(k+1)}\} \cup \{x,y,u',u,v',v\}$

$A(D_{uv})=\{c_ic_{i+1}:1\leq i \leq k(k+1) \} \cup
\{c_{2k}u',u'u,c_{k(k+1)-1}v',v'v\} \cup \{xy,xc_1,yc_1\}$

where addition is taken modulo $k(k+1)$.

Observe that in any homomorphism $f$ of $D_{uv}$ to $H_1$, we should
have $f(c_1)=1$. Once we assign the first $k$ vertices $c_1,\ldots
c_k$ color $1\ldots k$, the vertex $c_{k+1}$ is assigned with either
color 1 or color $k+1$. If we opt for color 1, then through the
whole remaining vertices $c_{k+1},\ldots ,c_{k(k+1)}$ we should
assign these vertices with colors along the $k-$cycle $12\ldots k$
in $H_1$. Else if we opt for color $k+1$, then we should assign the
whole remaining vertices with colors along the $(k+1)-$cycle
$12\ldots k+1$ in $H_1$. To see this, suppose to the contrary that
we assign the vertices $c_1,\ldots ,c_{k(k+1)}$ in $H$ with colors
along the $k-$cycle $s$ times and with colors along the
$(k+1)-$cycle $t$ times, where $0< t < k$. Then, we have the
following equation.

$k\cdot (k+1)= s\cdot k + t \cdot (k+1)$

which again implies

$(k+1)(k-t)=s \cdot k$

Knowing that the least common denominator of $k$ and $k+1$ is
$k(k+1)$, this leads to a contradiction. Hence, $(f(c_1),\ldots
,f(c_{k(k+1)}))$ coincides with one of the following sequences:

$(1,2,\ldots,k,\ldots ,1,\ldots,k)$: the sequence $1,2,\ldots ,k$
appears $k+1$ times. Or,

$(1,2,\ldots,k,k+1,\ldots ,1,\ldots,k+1)$: the sequence $1,2,\ldots
,k+1$ appears $k$ times.

If the first sequence is the actual one, then we have $f(c_{2k})=k$,
$f(u')\in \{1,k+1\}$, $f(u)\in \{1,2\}$, $f(c_{k(k+1)-1})=k-1$,
$f(v')=k$ and $f(v)\in \{1,k+1\}$. If the second one is the actual
one, then we have $f(c_{2k})=k-1$, $f(u')=k$, $f(u)\in \{1,k+1\}$,
$f(c_{k(k+1)-1})=k$, $f(v')\in \{1,k+1\}$ and $f(v)\in \{1,2\}$. In
both cases, we can assign both of $u$ and $v$ color 1. Furthermore
by choosing the right sequence, we can color one of $u$ and $v$ with
color 2 and the other with color 1. However we cannot assign color 2
to both $u$ and $v$ in a homomorphism.

Let $D$ be the digraph obtained by replacing every edge $uv \in
E(G)$ by $D_{uv}$. Here $D_{uv}$ is placed in an arbitrary
direction. Note that $|V(D)|=|V(G)|+|E(G)|\cdot (k(k+1)+4)$ and this
reduction can be done in polynomial time.

Let all costs $c_i(t)=0$ for $t\in V(D)$, $i\in V(H)$ apart from
$c_1(x)=1$ and $c_{k+1}(x)=|V(G)|$ for all $x \in V(G)$. Let $f$ be
a homomorphism of $D$ to $H$ and let $S=\{u\in V(G):f(u)=2\}$. Then,
$S$ is an independent set in $G$ since we cannot assign color 2 to
both $u$ and $v$ in $V(G)$ whenever there is an edge between them.
Observe that a minimum cost homomorphism will assign as many
vertices of $V(G)$ color 2.

Conversely, suppose we have an independent set $I$ of $G$. Then we
can build a homomorphism $f$ of $D$ to $H_1$ such that $f(u)=2$ for
all $u \in I$ and $f(u)=1$ for all $u \in G(V) \setminus I$. Note
that all the other vertices from $D_{uv}$, $uv \in E(G)$ can be
assigned with an appropriate color from $H_1$.

Hence, a minimum cost homomorphism $f$ of $D$ to $H_1$ yields a
maximum independent set of $G$ and vice versa, which completes the
proof.\qed

\begin{lemma}\label{h2}
Let $H'_2$ be a digraph obtained from $\overrightarrow{C_k}=12\ldots
k1, k \geq 3$, by adding an extra vertex $k+1$ such that $i \dom
k+1$ and $k+1 \dom i+1,i+2$, where $i, i+1, i+2$ are three
consecutive vertices in $\overrightarrow{C_k}$. Let $H_2$ be $H'_2$
or its converse. Then MinHom($H_2$) is NP-hard. (See Figure
\ref{H1andH2}.)
\end{lemma}

\pf Without loss of generality, we may assume that
$V(H_2)=\{1,\ldots ,k+1\}$, $123\ldots k$ is a cycle of length $k$,
and the vertex $k+1$ is dominated by $k$ and dominates 1 and 2.

We will construct a polynomial time reduction from the maximum
independent set problem to MinHOM($H_2$). Let $G$ be an arbitrary
undirected graph. We replace every edge $uv \in E(G)$ by the digraph
$D_{uv}$ defined as in the proof of Lemma \ref{h1}.

Observe that in any homomorphism $f$ of $D_{uv}$ to $H_2$, we should
have $f(c_1)=1$. And also by the same argument discussed in the
proof of Lemma \ref{h1}, the vertices of $k(k+1)-$cycle in $D_{uv}$
should be assigned with either along $k-$cycles, $12\ldots k$ and
$(k+1)2\ldots k$, or the $(k+1)-$cycle $12\ldots k+1$ in $H_2$. If
the vertices of $k(k+1)-$cycle in $D_{uv}$ are assigned with
$k-$cycles in $H_2$, then we have $f(c_{2k})=k$, $f(u')\in
\{1,k+1\}$, $f(u)\in \{1,2\}$, $f(c_{k(k+1)-1})=k-1$, $f(v')=k$ and
$f(v)\in \{1,k+1\}$. If the vertices of $k(k+1)-$cycle in $D_{uv}$
are assigned with $(k+1)-$cycles in $H_2$, then we have
$f(c_{2k})=k-1$, $f(u')=k$, $f(u)\in \{1,k+1\}$,
$f(c_{k(k+1)-1})=k$, $f(v')\in \{1,k+1\}$ and $f(v)\in \{1,2\}$. In
both cases, we can assign both of $u$ and $v$ color 1. Furthermore
by choosing the right sequence, we can color one of $u$ and $v$ with
color 2 and the other with color 1. However we cannot assign color 2
to both $u$ and $v$ in a homomorphism.

Now it is easy to check that the same argument as that in the proof
of Lemma \ref{h1} applies, completing the proof.\qed

Let $\mathcal I$ denote the following decision problem: given a
graph $X$ and an integer $k$, decide whether or not $X$ contains an
independent set of $k$ vertices. We denote by ${\mathcal I}_3$ the
restriction of $\mathcal I$ to graphs with a given three-colouring.
In the following Lemmas, we give a polynomial time reductions from
${\mathcal I}_3$. The following lemma shows that MinHom($O_i$) is
NP-hard for $i=1,2,3,4$.

\begin{proposition} \cite{mincostungraph} \label{MaxIndepNPhard}
The problem $\mathcal I$ is NP-complete, even when restricted to
three-colourable graphs (with a given three-colouring). \qed
\end{proposition}

\begin{lemma}
Let $H'$ be an arbitrary digraph over vertex set $\{1,2,3,4,5,6\}$
such that $$\{13,14,15,16,24,25,26,36,45,46\} \subseteq A(H'),$$
$$A(H') \subseteq \{12,13,14,15,16,24,25,26,36,45,46,56\}.$$ Let $H$ be
$H'$ or its converse. Then MinHOM($H$) is NP-hard. (See Figure
\ref{O1}.)
\end{lemma}
\pf Let $X$ be a graph whose vertices are partitioned into
independent sets $U, V, W$, and let $k$ be a given integer. We
construct an instance of MinHOM($H$) as follows: the digraph $G$ is
obtained from $X$ by replacing each edge $uv$ of $X$ with $u \in U,
v \in V$ by an arc $uv$, replacing each edge $vw$ of $X$ with $v \in
V, w \in W$ by an arc $vw$, and replace each edge $uw$ of $X$ with
$u \in U, w \in W$ by an arc $um_{uw},n_{uw}m_{uw},n_{uw}w$, where
$m_{uw},n_{uw}$ are new vertices. Define cost function $c_2(u)=0$,
$c_1(u)=1$, $c_3(v)=0$, $c_4(v)=1$, $c_5(w)=0$, $c_6(w)=1$,
$c_3(m_{uw})=c_3(n_{uw})=-|V(X)|$, $c_i(m_{uw})=c_i(n_{uw})=|V(X)|$
for $i \ne 3$. Apart from these, set all cost to $|V(X)|$.

We now claim that $X$ has an independent set of size $k$ if and only
if $G$ admits a homomorphism to $H$ of cost $|V(X)|-k$. Let $I$ be
an independent set in $G$. We can  define a mapping $f : V(G)
\rightarrow V(H)$ as follows:

\begin{itemize}
\item $f(u)=2$ for $u \in U \cap I$, $f(u)=1$ for $u \in U-I$
\item $f(v)=3$ for $v \in V \cap I$, $f(v)=4$ for $v \in V-I$
\item $f(w)=5$ for $w \in W \cap I$, $f(w)=6$ for $w \in W-I$.\\

When $uw \in E(X)$:
\item If $f(u)=2, f(w)=6$ then set $f(m_{uw})=6, f(n_{uw})=3$.
\item If $f(u)=1$ and $f(w) \in \{5,6\}$ then set $f(m_{uw})=3, f(n_{uw})=1$,

\end{itemize}
One can verify that $f$ is a homomorphism from $G$ to $H$, with cost
$V|X|-k$.

Let $f$ be a homomorphism of $G$ to $H$ of cost $|V(X)|-k$. If
$k\leq 0$ then we are trivially done so assume that $k>0$. Note that
we cannot assigne color 3 to both $n_{uw}$ and $m_{uw}$
simultaneously due to the arc $n_{uw}m_{uw}$. Hence, that the cost
of homomorphism $f$ is $|V(X)|-k$, $k>0$ implies that all the
vertices in $V(X)$ are assigned so that their individual costs are
either zero or one, and for each edge $uw\in E(X)$ the costs of
assigning $m_{uw}$ and $n_{uw}$ to vertices of $V(H)$ sum up to
zero.

Let $I=\{u \in V(X) \mbox{ $|$ } c_{f(u)}(u)=0 \}$ and note that
$|I|=k$. It can be seen that $I$ is an independent set in $G$, as if
$uw \in E(G)$, where $u \in I \cap U$ and $w \in I \cap W$ then
$f(u)=2$ and $f(w)=5$, which implies that $f(m_{uw}) \ne 3$ and
$f(n_{uw})\ne 3$ contrary to $f$ being a homomorphism of cost
$|V(X)|-k$.  \qed

\begin{lemma}\label{NPlocalysemicomplete}
Let $H$ be a connected locally semicomplete digraph which is neither
acyclic nor a directed cycle. Then MinHom($H$) is NP-hard.
\end{lemma}

\pf As $H$ is neither acyclic nor a directed cycle, it has an
induced cycle $\overrightarrow{C_k}=12\ldots k1, k \geq 2$ and a
vertex $k+1$ outside this cycle. For $H$ is connected, the vertex
$k+1$ is adjacent with at least one of the $\overrightarrow{C_k}$
vertices.

If $\overrightarrow{C_k}=\overrightarrow{C_2}$ and vertex $k+1$ is
adjacent with $1$, then $k+1$ is adjacent with vertex $2$ as well.
By Lemma \ref{NPsemicomplete} and Lemma \ref{subdigraph},
MinHom($H$) is NP-hard in this case.

Therefore, we assume that $H$ does not have any symmetric arc
hereinafter. Observe that the vertex $k+1$ cannot be adjacent with
more than four vertices of $\overrightarrow{C_k}$, since otherwise
$k+1$ either dominates or is dominated by at least three vertices on
$\overrightarrow{C_k}$, which is a contradiction by the existence of
a chord between two $\overrightarrow{C_k}$ vertices. With the same
argument, vertices which dominate or are dominated by $k+1$ are
consecutive on the cycle $\overrightarrow{C_k}$ and the number of
these vertices are at most two, respectively.

Now without loss of generality, assume that $k+1$ is dominated by
$1$ and is not dominated by $k$. Since both $k+1$ and $2$ are
outneighbors of vertex 1, there is an arc between $k+1$ and $2$.
Consider the following cases.

Case 1. $k+1 \dom 2$: The vertex $k+1$ either dominate $3$ or is
nonadjacent with $3$. Since $k+1$ is dominated by $1$, $k+1$ cannot
be dominated by $3$.

Case 1-1. $k+1 \dom 3$: The digraph $H[\{1,2,\ldots,k+1\}]$ is
isomorphic to $H_2$. Hence, MinHom($H$) is NP-hard by Lemma \ref{h2}
and Lemma \ref{subdigraph}. Observe that there is no arc between
$k+1$ and the vertices of $\overrightarrow{C_k}$ other than $1,2$
and $3$.

Case 1-2. $k+1$ is nonadjacent with $3$: There is no arc between
$k+1$ and the vertices of $\overrightarrow{C_k}$ other than $1$ and
$2$, thus MinHom($H$) is NP-hard by Lemma \ref{h1} and Lemma
\ref{subdigraph}.

Case 2. $2 \dom k+1$: Since $k+1$ and $3$ are outneighbors of vertex
2, there is an arc between $k+1$ and 3. Moreover, $k+1$ is dominated
by two vertices 1 and 2, which implies that $k+1 \dom 3$. Now the
vertex $k+1$ either dominate $4$ or is nonadjacent with $4$.

Case 2-1. $k+1 \dom 4$: The digraph $H[\{1,3,\ldots,k+1\}]$ is
isomorphic to $H_1$, thus MinHom($H$) is NP-hard by Lemma \ref{h1}
and Lemma \ref{subdigraph}.

Case 2-2. $k+1$ is nonadjacent with $4$: Observe that there is no
arc between $k+1$ and the vertices of $\overrightarrow{C_k}$ other
than $1,2$ and $3$. Hence, the digraph $H[\{1,2,\ldots,k+1\}]$ is
isomorphic to $H_2$. MinHom($H$) is NP-hard by Lemma \ref{h2} and
Lemma \ref{subdigraph}.\qed

\begin{lemma}\label{NPquasi}
Let $H$ be a connected quasi-transitive digraph which is neither
acyclic nor $\overrightarrow{C_2}$ nor an extension of
$\overrightarrow{C_3}$. Then MinHom($H$) is NP-hard.
\end{lemma}

\pf We can easily observe that $H$ has an induced cycle
$\overrightarrow{C_k}=12\ldots k1, k \geq 2$. If it has an induced
cycle $\overrightarrow{C_2}$, then there is a vertex $k+1$ outside
this cycle which is adjacent with one of the vertices in
$\overrightarrow{C_2}$. Furthermore, the quasi-transitivity of $H$
enforces $k+1$ to be adjacent with both vertices in this cycle, and
the cycle $\overrightarrow{C_2}$ together with $k+1$ induce a
semicomplete digraph. By Lemma \ref{NPsemicomplete} and Lemma
\ref{subdigraph}, MinHom($H$) is NP-hard in this case. Therefore, we
assume that $H$ does not have any symmetric arc hereinafter.

Note that $H$ cannot have an induced cycle
$\overrightarrow{C_k}=12\ldots k1$ of length greater than 3.
Otherwise, by quasi-transitivity of $H$ a chord appears in the
cycle, a contradiction. Hence we may consider only
$\overrightarrow{C_3}$ as an induced cycle of $H$. Choose a maximal
induced subdigraph $H'$ of $H$ which is an extension of
$\overrightarrow{C_3}$ with partite sets $X_1, X_2$ and $X_3$.
Clearly such subdigraph $H'$ exists.

By assumption $H'\neq H$ and we have a vertex $x$ which is adjacent
with at least one vertex of $H'$. Without loss of generality,
suppose that $x \dom 1$, for some $1\in X_1$. As $H$ is
quasi-transitive, vertex $x$ should be adjacent with every vertex of
$X_2$. There are two possibilities.

Case 1. $x\dom 2$ for some $2\in X_2$. Then $x$ is adjacent with every vertex $3\in X_3$ 
due to quasi-transitivity. Consider the subdigraph induced by $x$, $1,2$ and a vertex of $X_3$. MinHOM($H$) is NP-hard by Lemmas \ref{NPsemicomplete} and \ref{subdigraph}.

Case 2. $X_2\dom x$. Then there is an arc between $x$ and each
vertex of $X_1$ by quasi-transitivity. If $1'\dom x$ for some $1'\in
X_1$, $x$ is adjacent with every vertex of $X_3$ and MinHOM($H$) is NP-hard by Lemmas \ref{NPsemicomplete} and \ref{subdigraph}. Else if $x\dom X_1$, there is a
vertex $3\in X_3$ which is adjacent with $x$ since otherwise,
$H'\cup \{x\}$ is an extension of $\overrightarrow{C_3}$, a
contradiction to the maximality assumption. Again MinHOM($H$) is
NP-hard by Lemmas \ref{NPsemicomplete} and \ref{subdigraph}.\qed

{\small

\end{document}